\documentclass[11pt]{article}      

\evensidemargin=\oddsidemargin

\newcommand{\be}{\begin{equation}}
\newcommand{\ee}{\end{equation}}
\newcommand{\bea}{\begin{eqnarray}}
\newcommand{\eea}{\end{eqnarray}}
\newcommand{\p}{\partial}
\newcommand{\nn}{\nonumber}
\newcommand{\wt}{\widetilde}

\newcommand{\mH}{\mathbb H}

\newcommand{\Tr}{\operatorname{Tr}}

\def\half{\frac{1}{2}}
\usepackage{amsfonts,ulem,bbold}
\usepackage{amssymb}
\usepackage{amsmath}
\usepackage[usenames]{color}
\begin{document} 
\begin{titlepage}
\thispagestyle{empty}
{~}\\
\vskip 2 cm
\begin{center}
{\Large \bf  Exact Solution to the ``Auxiliary Extra   Dimension'' 
\\[.2cm]      
Model of Massive Gravity}      
\\
\vskip 0.7cm
{\bf S. F. Hassan}\footnote{e-mail: {\tt fawad@fysik.su.se}} 
{\bf and Rachel A. Rosen}\footnote{e-mail: {\tt rarosen@fysik.su.se}} 

\vspace{0.3cm} 

{\it  Department of Physics \& The Oskar Klein Centre, Stockholm
  University, \\[.1cm]
AlbaNova University Centre, SE-106 91 Stockholm, Sweden} 

\vskip 2cm

\begin{abstract}\noindent
The ``Auxiliary Extra Dimension'' model was proposed in order to
provide a geometrical interpretation to modifications of general
relativity, in particular to non-linear massive gravity. In this
context, the theory was shown to be ghost free to third order in
perturbations, in the decoupling limit. In this work, we exactly solve
the equation of motion in the extra dimension, to obtain a purely
4-dimensional theory. Using this solution, it is shown that the ghost
appears at the fourth order and beyond. We explore potential
modifications to address the ghost issue and find that their
consistent implementation requires going beyond the present framework.
\end{abstract}
\end{center}
\setcounter{footnote}{0}
\end{titlepage}
\newpage

\section{Introduction and Summary}

While general relativity is a geometric theory, the same cannot be
said of models of massive gravity. Moreover, it has been difficult to
construct such massive models that are free of ghost instabilities
\cite{FP1,FP2,BD}. The ``auxiliary extra dimension'' (AuXD) model was
proposed in an attempt to address both these issues simultaneously
\cite{G,dR}. In this model the mass term arises from the extrinsic
curvature of the $4$-dimensional spacetime in $5$ dimensions, the
$5^{th}$ dimension being non-dynamical. Thus in this model massive
gravity acquires a geometric interpretation. Subsequently, it was
verified in \cite{dRG1} that this theory was ghost free to $3^{rd}$
order in the ``decoupling limit''
\footnote{The decoupling limit corresponds to taking the graviton mass
  $m\rightarrow 0$ and $M_p\rightarrow \infty$ while keeping $m^2M_p$
  fixed. In practice, this means retaining terms to first order in the
  metric perturbation $h=g-\eta$, but to all orders in the non-linear
  St\"uckelberg fields $\pi$. Thus this limit allows one to consider
  the non-linearities that are most relevant to the ghost problem. It
  is mostly in this limit that the newly constructed theories of
  massive gravity are shown to be ghost free \cite{dRG2,dRGT}. In
  these models, the absence of the ghost to all orders away from the
  decoupling limit is still an open question (see, for example,
  \cite{ACM}). The ghost analysis of the AuXD model in the present
  paper is always done in the decoupling limit.\label{note1}}.

Since then, 4-dimensional theories of massive gravity that are
potentially ghost free to all orders have been constructed
\cite{dRG2,dRGT} and their systematics explored \cite{HR}. In fact,
the AuXD model seems to have been a motivation for revisiting the
earlier works \cite{AGS,CNPT} which led to the new developments. It is
therefore appropriate to determine where this model belongs in the
new scheme of things.

The AuXD model leads to two coupled equations: (i) a purely 4
dimensional Einstein-Hilbert equation with extrinsic curvature
contributions, and (ii) an equation for the extra dimension $u$. The
$u$-equation converts the extrinsic curvature contributions to mass
terms. So far, this equation has been solved perturbatively to third
order. It leads to a 4 dimensional massive gravity that is ghost free
to this order in the decoupling limit \cite{dRG1}. In the present
paper, we solve the $u$-equation {\it exactly} to obtain a purely
$4$-dimensional, closed-form expression for the non-linear mass term.
This allows us to compare the AuXD model to the recently constructed
theories of massive gravity \cite{dRG2,dRGT,HR} and examine its
stability to any order. We find that in the standard interpretation of
the model, the ghost re-enters at the $4^{th}$ order, and hence the
theory is not consistent. There exist non-standard modifications that
can potentially alleviate the ghost problem to any given order in the
decoupling limit. However, we show that such modifications cannot be
consistently implemented within the present setup.
(However, see footnote \ref{added}).    

The paper is organized as follows: The AuXD model is introduced in
section 2. In section 3 we solve the $u$-equation in terms of
integration constants. These are determined in section 4 in terms of
boundary conditions. There we obtain the 4-dimensional massive gravity
action and its equations of motion. In section $5$ this model is
compared to the potentially ghost free massive actions constructed
recently and a ghost is shown to appear at the fourth order and
beyond. We also explore modifications of the boundary condition in an
attempt to resolve the ghost issue.

\section{The Auxiliary Extra-dimension Model}

The starting point is the $4$-dimensional Einstein-Hilbert action with
an extrinsic curvature term involving the ``auxiliary
extra-dimension'' $u$ and $\wt g_{\mu\nu}(x,u)$ with $\mu,\nu =
0,1,2,3$ \cite{G,dR}, 
\be
S=-M_p^2\int d^4 x \left[\sqrt{g}R  + \frac{m^2}{2}
\int^{+1}_{-1}du\sqrt{\wt g}\,(k_{\mu\nu}k^{\mu\nu}-k^2)\right] 
+S_{\rm matter}.
\label{action}
\ee
Here, $k_{\mu\nu}=\frac{1}{2}\p_u\wt g_{\mu\nu}$, $k=\wt g^{\mu\nu}
k_{\mu\nu}$, and the 4-dimensional metric is $g_{\mu\nu}(x)=\wt
g_{\mu\nu}(x,0)$. The matter action $S_{\rm matter}$ is also localized
at $u=0$.  

To obtain the equations of motion, the action is varied with respect
to  $\wt g^{\mu\nu}(x,u)$. Then for $u\neq 0$ one obtains the
``$u$-equation'',  
\be
\label{ueq}
\frac{1}{\sqrt{\wt g}}\p_u\Big(\sqrt{\wt g}[k^{\mu\nu}-k\wt
  g^{\mu\nu}]\Big) -\frac{1}{2}\wt g^{\mu\nu}
(k_{\rho\sigma}k^{\rho\sigma}-k^2)
+2(k^\mu_{\,\,\lambda}k^{\lambda\nu}-k k^{\mu\nu})=0\,. 
\ee
Solving this requires specifying boundary conditions $\wt
g_{\mu\nu}(x,0)=g_{\mu\nu}(x)$ and $\wt g_{\mu\nu}(x,1)=f_{\mu\nu}(x)$,
for some $f_{\mu\nu}$. Furthermore, integrating the variation over
$u\in\{-\epsilon,\epsilon\}$ and assuming reflection symmetry about
$u=0$, one obtains the $4$-dimensional equation of motion,
\be
\label{4eq}
R_{\mu\nu}-\frac{1}{2}g_{\mu\nu}R + m^2(k_{\mu\nu}-
g_{\mu\nu}k)\Big|_{u=0^+} = G_N\, T_{\mu\nu}\,.
\ee
For $\wt g_{\mu \nu}(x,u)=\eta_{\mu \nu}+(1-u)h_{\mu \nu}(x)$, at the
linearized level, one recovers the Fierz-Pauli massive gravity
\cite{G,dR}. In \cite{dRG1} the model was analyzed to cubic order 
and was shown to be free of ghosts to that order. This supported
the proposal that the action (\ref{action}) provided a consistent
non-linear generalization of the Fierz-Pauli mass. The use of the
extra-dimension $u$ can then be understood as a way of packaging the
non-linearities. The flip side is that the non-linear structure of the 
mass term is not explicit in the model. Below, we solve for the 
$u$-dependence to obtain the purely $4$-dimensional form of the mass
term. Then, the massive theory can be written entirely in
$4$-dimensions with no reference to the extra-dimension. 

Before proceeding let us point out that the action (\ref{action}) is
invariant only under $4$ dimensional general coordinate
transformations that do not involve $u$. Being $u$-independent, these
transform $\wt g_{\mu\nu}(u=0)=g_{\mu\nu}$ and $\wt g_{\mu\nu}(u=1)
=f_{\mu\nu}$ in the same way. Hence $f_{\mu\nu}$ transforms as a rank
2 tensor and can be fixed to a specific form only in a given gauge.

\section{Solution of the $u$-Equation}

Using $\p_u\sqrt{\wt g}=\sqrt{\wt g}\,k$ and $k^\mu_{\,\,\nu}=\frac{1}{2}
\,\wt g^{\mu\sigma}\p_u\wt g_{\sigma\nu}$, the $u$-equation
(\ref{ueq}) becomes
\be
\p_u(k^\mu_{\,\,\nu}-k\delta^\mu_\nu)+k(k^\mu_{\,\,\nu}-k\delta^\mu_\nu)
-\frac{1}{2}\delta^\mu_\nu (k^\rho_{\,\,\sigma}k^\sigma_{\,\,\rho}-k^2)=0\,. 
\ee
We introduce the notation $\Bbbk$ for a matrix with elements
$k^\mu_{\,\,\nu}$ and denote its traceless part by $\Bbbk_t$ and its 
trace by $k$. In terms of these the $u$-equation splits into,
\bea
&\p_u \Bbbk_t = -k\, \Bbbk_t\,,&
\label{utraceless}\\
&\p_u k +\frac{1}{2} k^2 +\frac{2}{3}\, {\rm Tr}(\Bbbk_t^2)\,.&
\label{utrace}
\eea
Eliminating ${\rm Tr}(\Bbbk_t^2)$ from the above leads to a second
order equation $\p^2_u k + 3k\p_u k + k^3 =0$, with the solution,
\be
k(u)=\frac{2(u+c)}{(u+c)^2-d^2}=\p_u \ln[\,(u+c)^2-d^2\,]\,.
\label{solk}
\ee
The equation for $\Bbbk_t$ can now be integrated to,
\be
\Bbbk_t ={\mathbb C_t}\,[(u+c)^2-d^2]^{-1} \,.
\ee
Here, $c$, $d$ and the traceless matrix ${\mathbb C_t}$ are
integration constants to be determined in terms of the boundary
data $g_{\mu\nu}(x)$ and $f_{\mu\nu}(x)$. Demanding that these also
solve the first order equation (\ref{utrace}), determines $d$ as, 
\be
d^2=\tfrac{1}{3}\, {\rm Tr} ({\mathbb C}_t^2)\,.
\label{dCt}
\ee
The extrinsic curvature $k^\mu_{\,\,\nu}$ can now be reconstructed as 
(in matrix notation),
\be
\Bbbk = \Bbbk_t+\frac{1}{4}{\mathbb 1}\,k= 
\frac{{\mathbb C_t}+\frac{1}{2}(u+c){\mathbb 1}}{(u+c)^2-d^2}\,.
\label{solkmunu}
\ee
Now, the relation $k=\p_u(\ln\sqrt{\wt g})$ that follows from the
definition of $k_{\mu\nu}$ can be integrated, with the boundary
condition $\det\wt g(u=0)=\det g$, to give
\be
\sqrt{\wt g(u)}= \frac{(u+c)^2-d^2}{c^2-d^2}\sqrt{g} \, .
\label{detg}
\ee
These solutions can be used to perform the $u$-integral in
(\ref{action}) and obtain a purely $4$-dimensional action in terms of
the integration constants $c$ and $d$. Using the imposed $Z_2$
symmetry about $u=0$, and solutions (\ref{solk}), (\ref{solkmunu}) and
(\ref{detg}), the $u$-integral in (\ref{action}) becomes,
\be
I\equiv 2\int^{+1}_{0}du\sqrt{\wt g}(k_{\mu\nu}k^{\mu\nu}-k^2)= 
-\sqrt{g}\,\frac{6}{c^2-d^2}\,\int_0^1du \,.
\label{I}
\ee
Note that for the solution of the $u$-equation the integrand has
become completely $u$ independent! The metric equation of motion
(\ref{4eq}) in terms of integration constants, is,  
\be
R_{\mu\nu}-\frac{1}{2}g_{\mu\nu}R+m^2\frac{1}{c^2-d^2}
\left(g_{(\mu\lambda}\mathbb C^\lambda_{t\,\nu)}-\frac{3}{2}c\,
g_{\mu\nu}\right)=G_N\, T_{\mu\nu}\,.
\label{eom2}
\ee

\section{The 4-Dimensional Action}

We first determine the integration constants $c$, $d$, and 
$\mathbb C_t$ in terms of the boundary data $g_{\mu\nu}$ and
$f_{\mu\nu}$ at $u=0$ and $u=1$. Using the $u$-independent invertible
matrix $f_{\mu\nu}$ let us define, 
\be
E^\mu_{\,\,\nu}(u) = \wt g^{\mu\lambda}(u)\,f_{\lambda\nu} \, ,
\ee
or in matrix notation, $\mathbb E=\wt{\mathbb g}^{-1}
{\mathbb f}$. In terms of this variable we have, 
\be
\Bbbk(u)=-\frac{1}{2}\,\mathbb{E}^{-1}\p_u\mathbb{E}=
-\frac{1}{2}\,\p_u\ln \mathbb{E}(u) \,.
\label{lnE}
\ee
Of course, the last equality does not hold for a generic matrix
$\mathbb E(u)$. In the present case this is valid only because the
$u$-dependence of $\Bbbk$ in (\ref{solkmunu}) is not contained in its
matrix structure, but rather in the scalar coefficients of the
commuting matrices ${\mathbb C}_t$ and ${\mathbb 1}$. These will also 
  determine the matrix structure of $\mathbb E$ ensuring that
  $[\mathbb E\,,\p_u\mathbb E]=0$. $\mathbb E$ (with an upper and a
  lower index) is introduced so that power series expansions are
  defined unambiguously. Also, in the above construction,
  interchanging $f_{\mu\nu}$ and $g_{\mu\nu}$ will lead to the same
  eventual outcome. Integrating this, using the solution for $\Bbbk$
  (\ref{solkmunu}) gives,
\be
\ln\mathbb E(u)-\ln\mathbb E(0)=-2\int_0^u\Bbbk\,du=\mathbb D(u)-
\mathbb D(0) \, ,
\ee
where \footnote{Using $\int[(u+c)^2-d^2]^{-1}du=
-\tanh^{-1}(\frac{c+u}{d})/d+const.$ and $\tanh^{-1}(x)=\frac{1}{2}
\ln[(1+x)/(1-x)]$, which is valid for $|x|\leq 1$.},   
\be
\mathbb D(u)=\frac{\mathbb C_t}{d}\,\ln\left[\,\frac{d+c+u}{d-c-u}\,
\right] - \frac{\mathbb 1}{2} \ln\left[\,(u+c)^2-d^2\,\right]\, .
\ee
Specifically, for $u=1$, where, $\wt{\mathbb g}(1)=\mathbb f$, one has
$\ln\mathbb E(1)=0$, leading to,
\be
\ln(\mathbb g^{-1}\mathbb f)=\frac{\mathbb C_t}{d}\,
\ln\left[\,\frac{d-(c+1)}{d+c+1}\,\frac{d+c}{d-c}\,\right]
-\frac{\mathbb 1}{2} \ln\left[\,\frac{c^2-d^2}{(c+1)^2-d^2}\,\right]\, .
\ee
This determines the integration constants. The trace part gives (using
${\rm Tr}\ln\mathbb E=\ln\det\mathbb E$),  
\be
\sqrt{\det(\mathbb g^{-1}\mathbb f)}=\frac{(c+1)^2-d^2}{c^2-d^2} \, ,
\label{IC1}
\ee
which is equation (\ref{detg}) for $u=1$. To solve the traceless
equation, introduce $\mathbb L=\ln(\mathbb g^{-1}\mathbb f)$ and its
traceless part $\mathbb L_t$,
\be
\mathbb L_t=\ln(\mathbb g^{-1}\mathbb f)-\frac{\mathbb 1}{4}\,
{\rm Tr}\ln(\mathbb g^{-1}\mathbb f) \, .
\label{Lt}
\ee
Then, 
\be
\mathbb L_t=\frac{\mathbb C_t}{d}\,\ln\left[\,\frac{d-(c+1)}{d+c+1}\,
\frac{d+c}{d-c}\,\right]\,.
\ee
On squaring, tracing and using (\ref{dCt}) one gets, 
\be
e^{\sqrt{\frac{1}{3}{\rm Tr(\mathbb L_t^2)}}}
 = \frac{d-(c+1)}{d+c+1}\,\frac{d+c}{d-c} \, ,
 \label{IC2}
\ee
where, for later reference,
\be
{\rm Tr}(\mathbb L_t^2)={\rm Tr}\left[\ln(\mathbb g^{-1}\mathbb f)
\right]^2
-\frac{1}{4}\,
\left[{\rm Tr}\ln(\mathbb g^{-1}\mathbb f)\right]^2 \, .
\label{Lt2}
\ee
From these one finds that,
\be
\mathbb C_t=\frac{d\,\mathbb L_t}{\sqrt{\frac{1}{3}
\rm Tr(\mathbb L^2_t)}} \, . 
\ee
Multiplying and dividing (\ref{IC1}) and (\ref{IC2}) leads to, 
\be
\frac{1}{c\pm d}=\left[\det(\mathbb g^{-1}\mathbb f)\right]^{1/4}\,
e^{\mp \frac{1}{2}\sqrt{\frac{1}{3}{\rm Tr(\mathbb L_t^2)}}} -1 \, .
\ee

We can now use these expressions to write a $4$-dimensional action
and equation of motion entirely in terms of $g$ and $f$. For the
action, the relevant quantity is,
\be
\tfrac{1}{3} F(\mathbb g^{-1}\mathbb f) \equiv \frac{1}{c^2-d^2}
=\left[\det(\mathbb g^{-1}\mathbb f)\right]^{\frac{1}{2}} 
-2 \left[\det(\mathbb g^{-1}\mathbb f)\right]^{\frac{1}{4}} 
\cosh\left(\frac{1}{2\sqrt{3}}\sqrt{{\rm Tr}({\mathbb L}_t^2 )}
\right)+1 \,.
\ee
Then the non-linear action with the mass term (\ref{I}) becomes,
\be
S=-M^2_p\int d^4 x \sqrt{-g}\left[R(g)-m^2\,F(\mathbb g^{-1}\mathbb
  f) \right] + S_{\rm matter}\,.
\label{nlaction}
\ee
This has the generic structure of a non-linear massive gravity action.
Unsurprisingly, the boundary metric $f_{\mu\nu}$ has become the
auxiliary metric needed to formulate massive gravity (see, for
example, \cite{HR}). Note that this mass term can be written
equivalently in terms of $\mathbb f^{-1}\mathbb g$ with an appropriate
sign flip in the exponent of the determinants. The term ${\rm
  Tr}({\mathbb L}_t^2 )$ (\ref{Lt2}) will look the same either way.
The metric equation of motion (\ref{eom2}) can be fully expressed in
terms of $\mathbb g^{-1}\mathbb f$ using,
\be
\frac{1}{c-d}+\frac{1}{c+d}=\frac{2c}{c^2-d^2}\,,\qquad
\frac{1}{c-d}-\frac{1}{c+d}=\frac{2d}{c^2-d^2}\,.
\ee
One then obtains,
\bea
&R_{\mu\nu}-\frac{1}{2}g_{\mu\nu}R+m^2 \left[
[\det(\mathbb g^{-1}\mathbb f)]^{\frac{1}{4}}\sinh\left(\frac{1}{2\sqrt{3}}
\sqrt{{\rm Tr}({\mathbb L}_t^2 )}\right) \left[\tfrac{1}{3}{\rm
      Tr}({\mathbb L}_t^2 )\right]^{-\half} \right]  
g_{(\mu \lambda} {\mathbb L}^\lambda_{t\,\nu)} &
\nonumber \\  [.3cm]
&\qquad -\tfrac{3}{2}m^2 \left[[\det(\mathbb g^{-1}\mathbb f)]^{\frac{1}{4}}
  \cosh\left(\frac{1}{2\sqrt{3}} \sqrt{{\rm Tr}({\mathbb L}_t^2 )}
  \right)-1\right]g_{\mu \nu}=G_N\, T_{\mu\nu}\,. &
\label{nleom}
\eea
This equation can also be obtained directly by varying the
4-dimensional action (\ref{nlaction}) with respect to $g_{\mu \nu}$, 
for fixed $f_{\mu\nu}$.  

\section{The Status of the Ghost Problem}

The above solution is valid for any fixed $f_{\mu\nu}$. In order to
obtain the Fierz-Pauli Lagrangian for massive gravity at lowest order
in the fields, the standard approach is to take $f_{\mu\nu}$ to be
flat \cite{G,dR,dRG1}. With this premise, one can now verify that the
action (\ref{nlaction}) contains a Fierz-Pauli mass and check if it
can avoid the ghost instability. We show that for the standard
interpretation of $f_{\mu\nu}$ the theory is not ghost free. A
non-standard interpretation is also discussed below.

In the standard massive gravity context, $f_{\mu\nu}$ is the coordinate
transform of the flat metric,  
\be
f_{\mu\nu}=\eta_{ab}\p_\mu\phi^a \p_\nu\phi^b \, .  
\label{f}
\ee 
Let us introduce the $(1,1)$ tensor $H^\mu_{\,\,\nu}$ so that,
\be 
\mathbb g^{-1}{\mathbb f}=\mathbb 1-\mathbb H\,, 
\label{H}
\ee 
where as usual $\mH$ denotes the matrix with elements
$H^\mu_{\,\,\nu}$. In \cite{dRG1}, the most general, potentially
ghost free, massive action was written to quintic order as a
polynomial in $\mH$, and with free parameters $c_3$, $d_5$ --
 the expression that multiplies $f_7$ vanishes in 4 dimensions 
\cite{HR}. Now, expanding the mass term in (\ref{nlaction}) to
$5^{th}$ order in $\mH$ one obtains,    
\bea
F(g^{-1}f)&=&-\frac{1}{4}\bigg\{
\Big[-(\Tr\mH)^2+\Tr\mH^2\Big]+\frac{1}{4}\Big[(\Tr\mH)^3
  -5\Tr\mH\Tr\mH^2+4\Tr\mH^3\Big]\nn\\ 
&+&\frac{1}{12^2}\Big[-5(\Tr\mH)^4+58(\Tr\mH)^2\Tr\mH^2
-53(\Tr\mH^2)^2-132\Tr\mH\Tr\mH^3+132\Tr\mH^4\Big] \nn\\ 
&+&\frac{1}{24^2}\Big[2(\Tr\mH)^5-41(\Tr\mH)^3\Tr\mH^2
  +123\Tr\mH (\Tr\mH^2)^2+160(\Tr\mH)^2\Tr\mH^3
  \nn\\ 
&&\qquad -304\Tr\mH^2\Tr\mH^3 -420\Tr\mH
  \Tr\mH^4+480\Tr\mH^5)\Big]+O(\mH^6) \bigg\}\,.
\label{Fquintic}
\eea
The quadratic expression is the Fierz-Pauli mass. 
For the choice $c_3=\tfrac{1}{4}$, the cubic terms match with the
corresponding terms in \cite{dRG2} that are ghost free in the
decoupling limit, as was first shown in \cite{dRG1}. However, at
the quartic order and beyond, no value of $d_5$ in \cite{dRG2} can
reproduce the corresponding terms here. This implies that the AuXD
model is not ghost free beyond the cubic order\footnote{The presence
  of the ghost can also be expected on general grounds. As stressed in
  \cite{HR}, the potentially ghost free models of \cite{dRG2} have a
  more natural expression in terms of ${\mathbb g}^{-1}\mathbb f$,
  rather than ${\mathbb f}^{-1}\mathbb g$. This is not true of
  (\ref{nlaction}) which treats $g_{\mu\nu}$ and $g^{\mu\nu}$ more
  symmetrically, a property traced back to the structure of
  $k^\mu_{\,\,\nu}$.}$^,$\footnote{After the completion of this work,
  we were informed by C. de Rham and G. Gabadadze that the extension
  of their third order calculation \cite{dRG1} to fourth order leads
  to the same conclusion.}.
In fact, the closest ghost free expression
corresponds to $d_5=-5/12^2$ which gives the quartic term coefficients
$\{-5,57,-51, -130, 129\}$ rather than the $\{-5,58,-53, -132, 132\}$
found above. 

{\it Can the ghost problem be cured?} To identify the AuXD model with
massive gravity we required $f_{\mu\nu}$ to be flat.
This was consistent with the potentially ghost-free theory to cubic
order. One may consider more general $f_{\mu\nu}$ to attempt to
resolve the ghost issue.

In particular, one may regard $f_{\mu\nu}$ as an arbitrary function of
both $g_{\mu\nu}$ and the matrix given by the right hand side of
(\ref{f})\footnote{We would like to thank G. Gabadadze for suggesting
  this possibility.}. With this new interpretation of $f_{\mu\nu}$ we
can write,  
\be
{\mathbb g}^{-1}{\mathbb f}  = {\mathbb 1}- M(\mH) \, ,
\label{MH}
\ee
where $\mH$ is defined such that $H_{\mu\nu}=g_{\mu\nu}-\eta_{ab}
\p_\mu\phi^a \p_\nu\phi^b$. 
Now, one can take the equations of motion (\ref{nleom}) for such a  
choice of $f_{\mu\nu}$ and identify them with the equations of motion
for the potentially ghost-free actions, for example equation (4.20) in
ref. \cite{HR} where the auxiliary metric there is taken to be flat.
We can use this as a definition of $\mathbb f$, or equivalently of
$M$, in (\ref{MH}). However, while such a procedure guarantees that
the equation of motion has the correct ghost-free structure (at least
in the sense of \cite{AGS,CNPT,dRG2}), the resulting equation can no
longer be regarded as the equation of motion for the AuXD model. This
is because the boundary metric $\mathbb f$ is now a function of
$\mathbb g$ while the equations of motion were derived assuming 
that $\wt{\mathbb g}$ did not vary on the boundary.

In other words, deriving the $u$-equation (\ref{ueq}) from
the the AuXD action (\ref{action}) required setting to zero a boundary
term,
\be
\sqrt{\wt g}\,\left(k^\mu_{\,\nu}-k\delta^\mu_\nu\right)\,
\wt g^{\nu\lambda}\,\delta \wt g_{\lambda\mu}\big\vert_{u_0}\,.
\label{b}
\ee
In the standard interpretation, the boundary is at $u_0=\pm 1$, and
there, $\wt g_{\mu \nu}(u=\pm 1)=f_{\mu \nu}$. For an $f_{\mu \nu}$
independent of $g_{\mu\nu}$, this is consistent with $\delta\wt g=0$.
However, if $f_{\mu \nu}$ depends on $g_{\mu\nu}$, then this 
variation does not vanish. From our solution it is also clear that the 
coefficient of $\delta\wt g$ does not vanish at $u=1$. Thus the
boundary term (\ref{b}) is not zero.

Another option is to derive the $u$-equation for $u_0=\pm
\infty$, assuming that the boundary term vanishes there, but find a
solution subject to the boundary condition at $u=1$. However, even in
this case, our solutions determine the behavior of $\wt g$ beyond
$u=1$. In fact, from (\ref{lnE}), $\ln{\mathbb E}={\mathbb D}(u)-
{\mathbb D}(1)$ and then it is not difficult to see that,
\be
\lim_{u\rightarrow\infty}\wt g_{\mu\nu}(u)=u \,A_{\mu\nu}(x)\,,
\ee
Here $A_{\mu\nu}$ is a $u$-independent matrix that depends on $g_{\mu
  \nu}(x)$. Thus, in this limit, a variation $\delta g$ at $u=0$
induces a variation $\delta\wt g$ at $u=\infty$. Moreover, in this
limit, $k^\mu_{\,\nu}=u^{-1} \delta^\mu_\nu$ and therefore, the
coefficient of $\delta\wt g$ in the boundary term becomes a
$u$-independent non-zero factor. Hence we see that this non-standard
interpretation of $f_{\mu\nu}$ is not consistent with the variation
principle that gives the $u$-equation. To implement such a
non-standard interpretation of $f$ in AuXD model, one will have to
modify the action to make it consistent with the variation
principle \footnote{However, it was pointed out more
    recently in \cite{BM} that the boundary term (\ref{b}) modifies
    only eqn. (\ref{4eq}) and not the $u$-equation (\ref{ueq}). This
    keeps our solution (\ref{nlaction}) unchanged. In this way,
    \cite{BM} was able to tune the mass term order by
    order. \label{added}}.   

It is also possible that adding specific higher order curvature terms,
like $K_{\mu \nu}^3$, to the action can address the ghost issue.
However, all these approaches to the problem require foreknowledge of
the ghost-free massive gravity action, contrary to the initial
approach of \cite{G,dR} in which the ghost-free structure emerged
naturally to cubic order. It is not obvious that such modifications,
even if implemented consistently, would still preserve the geometric
interpretation of the mass term which was a virtue of the original
AuXD model of \cite{G,dR}.

\vspace{0.3cm}

\begin{center}
{\bf   Acknowledgments}
\end{center}

We would like to thank Claudia de Rham, Gregory Gabadadze and Justin
Khoury for useful discussions. The work of RAR is supported by the
Swedish Research Council (VR) through the Oskar Klein Centre.

\end{document}